\documentstyle[preprint,prl,aps]{revtex}
\begin{document}
\draft
\title{Magnetic-Field Induced First-Order Transition in the
        Frustrated XY Model on a Stacked Triangular Lattice}
\author{M.L. Plumer, A. Mailhot,\cite{note} and A. Caill\'e}
\address{ Centre de Recherche en Physique du Solide et D\'epartement de
Physique}
\address{Universit\'e de Sherbrooke, Sherbrooke, Qu\'ebec, Canada J1K 2R1}
\date{March 1993}
\maketitle
\begin{abstract}
The results of extensive Monte Carlo simulations of magnetic-field
induced transitions in the xy model on a stacked triangular lattice
with antiferromagnetic intraplane and ferromagnetic interplane
interactions are discussed.  A low-field transition from the
paramagnetic to a 3-state (Potts) phase is found to be very weakly
first order with behavior suggesting tricriticality at zero field.
In addition to clarifying some long-standing ambiguity
concerning the nature of this Potts-like transition, the present work
also serves to further our understanding of the critical behavior at
$T_N$, about which there has been much controversy.
\end{abstract}

The possibility of novel critical behavior associated with geometrically
frustrated antiferromagnets has given rise to a wide variety of speculation
\cite{plumA}.
In the cases of Heisenberg and xy models on the stacked triangular lattice,
Kawamura \cite{kawaA} has argued, by means of symmetry analysis and a
$4-\epsilon$
renormalization group expansion, in favor of a new chiral universality
class with unusual critical exponents as determined by Monte Carlo simulations.
A field theoretic $2+\epsilon$ expansion by
Azaria, Delamotte and Jolicoeur \cite{aza}, however, has inspired the
suggestion
that such systems exhibit nonuniversality where
first-order, mean-field tricritical or O(4) criticality can occur depending
on unspecified details of the model
(also see Ref.\cite{chub}).
Two very recent Monte Carlo simulations on the Heisenberg model have
yielded different conclusions dependent upon details of the analysis.
Bhattacharya {\it et al} \cite{bhatt} maintain
that this system exhibits O(4) universality whereas our own work \cite{mailA}
supports
the idea of a very weak first-order transition with possibly a pseudo-critical
region \cite{pecz} giving O(4) critical exponents.
Two similar numerical studies of the Ising model, although yielding
substantially the same results, have led to the speculation of yet
another new universality class by one group \cite{bunk},
in contrast with our intrepretation
\cite{plumB} that the previous suggestion of standard xy universality
\cite{blank} is confirmed.  It is also of interest to note that the Monte Carlo
results which led to the proposal of a new
universality class associated with the frustrated pyrochlore antiferromagnet
\cite{reim} have been re-interpreted in support of a first-order
transition \cite{mailB}.  It is becoming increasingly clear that the results
of direct numerical simulations of frustrated spin systems can be
difficult to interpret.  The present work represents an attempt to
reveal the critical behavior of the xy model on a stacked triangular lattice
by application of a magnetic field H.  The analysis of these Monte-Carlo
simulation results is guided by expected behavior based on symmetry arguments
of a less controversial nature and lend
support to the proposal of tricriticality (or an extremely weak first-order
transition) for this system.

This work was inspired by the extensive study of Lee {\it et al} \cite{lee}
who examined the xy antiferromagnet on a triangular lattice (un-stacked) in an
applied magnetic field.  At H=0, the transition exhibits Kosterlitz-Thouless
behavior but the field breaks rotational symmetry and
transitions involving true long-range spin order occur.
For $H>0$, but not too large,
a colinear phase is stabilized with the {\it symmetry} of the 3-state Potts
model.  The real 3-state Potts model exhibits a continuous transition in
two dimensions (2D) and, after numerous numerical simulations over
the past twenty years,
it appears to be generally accepted that the transition is weakly first
order for the 3D lattice \cite{wu,potts1}.
For the so-called continuous 3-state Potts model, an effective
Landau-Ginsburg-Wilson (LGW) Hamiltonian is constructed which contains a
term cubic
in the order parameter \cite{steph}. Within mean-field theory, such models
yield a first-order transition, independent of space dimensionality.
The transition in 2D is
thus driven to be continuous by critical fluctuations, with known critical
exponents verified by the work of Lee {\it et al}.  Some analyses of
renormalization-group and series expansions for the 3D case indicate
a transition which may be continuous but most studies
favor the first-order scenario \cite{wu,ham}.  The conclusion of an earlier
Landau-type
analysis, which partially included effects of fluctuations, is that
models of this type may exhibit either a continuous or first-order
transition depending
on relative parameter values \cite{alex}.  Although contrary results
would have been surprising, it was not {\it a priori} certain that the
transition to the 3-state ordered phase considered here would be first order.
We find convincing evidence that the transition is indeed weakly first order.
This study represents the first Monte Carlo simulation of a model equivalent
to the continuous 3-state Potts Hamiltonian in 3D.

We study the Hamiltonian
\begin{equation}
{\cal H}~=~J_{\|} \sum_{<ij>} {\bf S}_i \cdot {\bf S}_j
\label{eq1}
\end{equation}
where the spins lie in the basal plane, $J_\|$ is the interplane interaction,
$J_{\bot}>0$ indicates
the antiferromagnetic coupling which is frustrated for the triangular
geometry, $<i,j>$
and $<k,l>$ represent near-neighbor sums along the hexagonal $c$ axis
and in the basal plane, respectively, and the field is applied
in the basal plane direction $x$.  The magnetic order realized by
this model can be conveniently described in terms of a spin density
expressed as a low-order Fourier expansion \cite{plumA}
\begin{equation}
{\bf s(r)}~=~{\bf m} + {\bf S}e^{i{\bf Q \cdot r}}
                     + {\bf S^{\ast}}e^{-i{\bf Q \cdot r}}
\label{eq2}
\end{equation}
where {\bf m} is the uniform component induced by the magnetic field,
{\bf Q} is the wave vector, and
the complex polarization vector can be written in terms of real
vectors, ${\bf S} = {\bf S}_a + i{\bf S}_b$.
The $120^\circ$ spin structure known
to occur at zero field below the N\'eel temperature $T_N$ is described
by a period-3 basal-plane modulation, along with a helical
polarization for {\bf S}.  At H=0,
the critical behavior is independent of the sign of the interaction $J_\|$.
For $J_\|<0$ there is no interplane modulation but for $J_\|>0$ a
period-2 structure is stabilized.
This difference gives rise to a term cubic in {\bf S} in a Landau-type
free energy, or LGW Hamiltonian, {\it only} for the case $J_\|<0$:
$F_3 \sim ({\bf m}\cdot{\bf S}){\bf S}\cdot{\bf S}) + c.c.$.  For a linearly
polarized spin density, one can write ${\bf S}={\bf S}_re^{i\phi}$,
where ${\bf S}_r\|{\bf m}$ is real, to get a contribution
$F_3 \sim mS_r^3cos(3\phi)$.
The three (Potts) states arise from the three inequivalent choices of
the phase angle $\phi=n\pi/3$, where n is an integer.
Higher-order terms in the free energy can stabilize $\phi=(2n+1)\pi/6$
depending on field and temperature values \cite{plumC}. In this case the
cubic term is zero.

In order to determine the magnetic-field temperature phase diagram,
standard Monte Carlo simulations were performed on the Hamiltonian (1)
with $J_\|=-1$ and $J_\bot=1$
for lattices $L \times L \times L$
with L=12-24.  Runs of $1-2\times10^4$ Monte Carlo steps (MCS) per spin
were made, with the initial $4-8\times10^3$ MCS discarded for thermalization.
Boundary lines were estimated as in
our previous work \cite{plumD}.  Not surprisingly, the result shown
in Fig. 1 is similar to the 2D case studied by Lee {\it et al}.
In particular, in addition to the paramagnetic phase 1,
there are two ordered phases, 6 and 9, with a linear
polarization of the spin density and an elliptical phase 7
(phases are numbered following a previous convention \cite{plumE}).
At H=0, the expected $120^\circ$ spin configuration was observed, with
the N\'eel temperature $T_N \simeq 1.45$ in agreement
with Kawamura \cite{kawaA} and Ref. \cite{plumD}.
These phases have the same symmetry as determined in the 2D case.
Phase 6 with  ${\bf S}\|{\bf H}$ is the 3-state Potts phase
discussed by Lee {\it et al}.

A molecular field treatment of this model \cite{leeB}, which yields
results independent of dimension, gives a phase diagram where the linear state
6 is absent and two critical lines which merge at $T_N$.  This is somewhat
puzzling since the cubic term responsible for the stability of phase 6 occurs
in the expansion of the free energy, in powers of {\bf s}({\bf r}), which
can be derived within this mean-field theory.
Such an expansion gives an expression for F({\bf m},{\bf S}) which is
identical in structure
to what is obtained for a phenomenological Landau-type free energy
(also mean-field) \cite{plumC,plumD}.  The only difference is that in the
latter case
each of the (six) fourth-order terms have an independent coefficient
$B_i$ (since each term is an independent invariant), whereas in the
molecular-field treatment all these coefficients are equal.  We
find that by making just one of the $B_i$ different from the
others, a phase diagram with the correct structure (Fig. 1) is found,
where the 1-6 boundary is first order as expected \cite{plumF}.
Molecular-field theory appears to somewhat accidentaly exclude phase 6.
This model also yields the result that the 1-9 phase boundary represents
a line of continuous transitions since the phase angle $\phi$
approaches the value
$\pi/6$ at this boundary line so the cubic term is not relevant.

The criticality of the 1-6 transition boundary was
studied at two points using
the Ferrenberg-Swendsen histogram method \cite{ferrA} of analysing Monte
Carlo data.  This technique is well-suited for the study of
transitions which may be very weakly first order, particularly when used
to determine the internal-energy cumulant \cite{reim,challa}
\begin{equation}
U(T)~=~1 - \frac13<E^4>/<E^2>^2.
\label{eq3}
\end{equation}
This quantity exhibits a minima near $T_N$, which achieves the value
$U^\ast = \frac23$ in the limit $L \rightarrow \infty$ for continuous
phase transitions.  In the case of a first-order transition, $U^\ast<\frac23$
is expected.  The histogram method may also be used to determine precisely
the location of extrema near $T_N$ which occur in other thermodynamic
functions. These are expected to demonstrate simple asymptotic volume
dependence in the case of first-order transitions or have L-dependence
governed by critical-exponent ratios in the case of continuous transitions.
In addition to $U$, results are given here for
the specific heat ($C$), staggered susceptibility
($\chi$) as well as the logarithmic derivative of the order parameter ($V$)
which is equivalent to \cite{ferrB}
\begin{equation}
V(T)~=~<ME>/<M> - <E>,
\label{eq4}
\end{equation}
where the relevant order parameter $M$ is defined as in Ref.\cite{plumD}.
Simulations were performed at H=0.7 and H=1.5 on lattices with L=12-33.
Thermodynamic averages were made using $1 \times 10^6$ MCS for the smaller
lattices and up to $2.6 \times 10^6$ MCS for the larger lattices, after
discarding the initial $2-3 \times 10^5$ MCS for thermalization. If
necessary, several runs at different T were made to ensure
that the extrema of the desired function occured close to at least one
simulation temperature.  The estimated critical temperatures are
1.488(2) and 1.522(2) for H=0.7 and 1.5, respectively.

Asymptotic scaling of the extrema with volume for the case H=0.7
demonstrated in Figs. 2-4 is consistent with a
first order transition \cite{pecz}.  An indication that it is only
weakly first order
is revealed by noting that the estimate $U^\ast=0.66660(3)$
from Fig. 2
is more than an order of magnitude closer to $\frac23$ than the value
determined for the 5-state Potts model in 2D \cite{fuk}, considered to be one
of the weakest first-order transitions known.  It is also of interest to
note that the value $U^\ast=0.6460(2)$ was determined by Fukugita {\it et al}
\cite {potts1} for the discrete 3-state Potts model on a cubic lattice,
also known to be only weakly first order \cite{bunk2}.
Similar first-order bevavior
was found in the data at H=1.5.  Evidence that the 1-6 transition becomes
more strongly first order as H increases is given by the estimate
$U^\ast=0.66643(3)$ at this higher field value.  Histogram data were also
taken at H=0.7 for the 6-7 transition, which has a critical temperature
estimated to be 1.420(2).  In a manner similar to our
analysis of a continuous transition on the stacked triangular lattice
\cite{plumB}, scaling consistent with the Ising universality class is evident.
This is the expected result as this transition involves only the order
parameter $S_{by}$.  Details of these results, along with histogram
analyses of the other two transition lines, will be presented elsewhere
\cite{plumF}.  Preliminary results for the 1-9 transition indicate this
transition is continuous, as it is within the phenomenological Landau-type
model discussed above.

The corresponding phase diagram for the case of antiferromagnetic
interplane coupling has only one linear state
with ${\bf S}\bot{\bf H}$ and two critical
lines emanating from $T_N$, which are transitions of xy ($S_1$) and Ising
($Z_2$) universality \cite{plumD}.  This structure nicely reveals the
$Z_2 \times S_1$ symmetry of the order parameter for the transition at
$T_N$ \cite{kawaA}.  The results of the present study suggest that this
picture does not occur in the case of $J_\|<0$ (or for $J_\|>0$ with
an applied field staggered along the $c$ axis).

The Monte-Carlo histogram simulations of this work
give clear indication that the present version of the continuous
3-state Potts model exhibits a very weak first-order transition in 3D.
In addition, our results suggest that this
transition becomes more weakly first order as
the field is lowered.  $T_N$ thus appears to have characteristics in common
with a tricritical point, a possibility suggested by Azaria {\it et al}
\cite{aza}.  This scenario is made somewhat ambiguous by the fact
that $T_N$ is also a multicritical point where more than one phase
meet.  The conventional Monte Carlo simulations of Kawamura \cite{kawaA}
were used to estimate critical exponents associated with his
proposed chiral universality class which are not very different from
those expected for mean-field tricriticality (a possibility that
was considered in Ref.\cite{kawaA}).  In view of these results and
the recent ambiguity
in interpreting Monte Carlo data for the frustrated Heisenberg model
\cite{bhatt,mailA},
as well as the discussion by Peczak and Landau \cite{pecz} of
pseudo-critical behavior associated with weakly first-order transitions,
our results are consistent with the transition at $T_N$ for the xy model
being tricritical
or very weakly first order.  It appears that only histogram Monte-Carlo
simulations at $T_N$ which are very extensive (long runs on large
lattices) have the possibility to add new information on this problem.

\acknowledgements
We thank P. Azaria, A.N. Berker, A. Chubukov, B. Delamotte, H.T. Diep,
and S. Miyashita for useful discussions. This work was supported by NSERC of
Canada and FCAR du Qu\'ebec.
%

\begin{figure}
\caption{Phase diagram determined by standard Monte Carlo simulations (points
with error bars).  Indicated are the paramagnetic
phase 1, phases 6 and 9 having colinear order and phase
7 with an elliptical (chiral) spin structure.  Squares at H=0.7 and 1.5
indicate
boundary points determined by highly accurate histogram analyses.
Solid and broken lines are guides to the eye and
indicate first and second-order transitions, respectively.}
\label{fig1}
\end{figure}

\begin{figure}
\caption{Scaling of the energy-cumulant minima with volume
for the 1-6 transition at H=0.7  The straight line represents a fit
to the four largest lattice sizes.}
\label{fig2}
\end{figure}

\begin{figure}
\caption{Scaling behavior of the specific heat maxima with volume
as in Fig. 2.}
\label{fig3}
\end{figure}

\begin{figure}
\caption{Scaling behavior of the maxima of the susceptibility $\chi$
and logarithimic derivative of the order parameter $V$ as in
Fig. 2.}
\label{fig4}
\end{figure}

\end{document}